\documentclass[runningheads]{llncs}
\usepackage[T1]{fontenc}
\usepackage{graphicx}
\usepackage{tabularx} 

\begin{document}
\title{Contribution Title}
%
%
\title{CPU-Based Layout Design for Picker-to-Parts Pallet Warehouses}
\author{Timo Looms\inst{1} \and
Lin Xie\inst{2}}
\authorrunning{Looms and Xie}
%
\institute{University of Twente, Netherlands \and
Brandenburg University of Technology Cottbus-Senftenberg, D-03046 Cottbus, Germany\\
\email{lin.xie@b-tu.de}}
\maketitle              
\begin{abstract}
Picker-to-parts pallet warehouses often face inefficiencies due to conventional layouts causing excessive travel distances and high labor requirements. This study introduces a novel layout design inspired by CPU architecture, partitioning warehouse space into specialized zones— Performance (P), Efficiency (E), and Shared (S). Discrete-event simulation is used to evaluate this design against traditional rectangular (random and ABC storage) and Flying-V layouts. Results demonstrate significant improvements in throughput time and reduced labor requirements, highlighting the potential for CPU-based layouts in optimizing warehouse operations.

\keywords{Warehouse Layout \and CPU-Inspired Design \and Picker-to-Parts \and Simulation.}
\end{abstract}
\section{Introduction}
Order picking is critical to warehouse operations, often accounting for more than half of operational costs. Effective layout design significantly impacts picking efficiency by influencing picker travel distances, order accuracy, and operational throughput. An optimized layout minimizes unnecessary movement and improves retrieval speeds, thereby directly reducing labor costs and enhancing productivity \cite{DeKoster2007}.

Picker-to-parts systems frequently struggle with excessive picker travel, inefficient inventory allocation, and suboptimal space utilization. Traditional orthogonal warehouse layouts are prevalent due to their simplicity but are inefficient in minimizing travel distances. Non-orthogonal layouts, such as Fishbone and Flying-V, aim to reduce picker travel, albeit at the expense of increased complexity and spatial inefficiency \cite{Cardona2012},\cite{Gue2009}.

To address these issues, we introduce a CPU-inspired layout design that segments warehouse space into Performance (P), Efficiency (E), and Shared (S) zones, optimizing labor efficiency and reducing travel distances. This study aims to evaluate the effectiveness of this novel layout through detailed discrete-event simulation.

\section{Related Work}
\label{sec:related_work}

Warehouse layout design spans multiple decision levels. At the \textit{macro level}, designers address compartment zoning, equipment selection, and the overall warehouse shape (e.g., U-, I-, or L-shaped). The \textit{meso level} focuses on geometric layout—such as Fishbone, Flying-V, or the proposed CPU-based design—while the \textit{micro level} concerns aisle width, slotting logic, and picker routing strategies \cite{baker2009warehouse}. Related discussions can also be found in \cite{Masae2024Warehouse} and \cite{roodbergen2009warehouse}. This work focuses on the meso level, where layout geometry interacts closely with storage policies and operational efficiency.

Early studies, such as Heragu et al.~\cite{Heragu2005Mathematical} and Hwang and Cho~\cite{HwangCho2006}, applied analytical and simulation models to evaluate layout–space trade-offs. More recent surveys, including those by Saini et al.~\cite{Saini2025Simulation} and Zaerpour et al.~\cite{Zaerpour_deKoster_Yu_2013}, emphasize the need for integrated performance evaluation methods—ranging from discrete-event simulation (DES) to queuing-based models.

A variety of geometric layouts have been proposed beyond the traditional grid-based form. The Flying-V and Fishbone layouts, analyzed extensively by Pohl et al.~\cite{Pohl2011Turnover} and Esmero et al.~\cite{Esmero2021Heuristic}, demonstrate potential gains in picker travel distance under randomized allocation. Further analyses can be found in \cite{Cardona2012Analytical}, \cite{Cardona2015Detailed}, and \cite{Dukic2008Analysis}. Petersen et al.~\cite{Petersen2004Improving} compared class-based and random storage using DES.

Methodologically, discrete-event simulation remains the dominant tool for analyzing warehouse layout and slotting-policy interactions. Macro and Salmi~\cite{Macro2002Simulation} applied simulation to evaluate warehouse efficiencies and storage allocations, establishing its value in layout assessment. Yener and Yazgan~\cite{Yener2019Optimal} extended this approach by integrating simulation with analytical modeling to optimize warehouse design in a real-world case study. Derhami et al.~\cite{Derhami2020Simulation} employed simulation-based optimization to determine efficient block-stacking layouts, while Esmero et al.~\cite{Esmero2021Heuristic} used heuristic simulations to compare non-conventional configurations. Collectively, these studies highlight the effectiveness of simulation in evaluating and improving warehouse design strategies.

Recent research indicates a shift in warehouse layout design from traditional homogeneous storage toward systems that accommodate heterogeneous products. According to Albert et al.~\cite{Albert2023Trends}, modern warehouses increasingly manage items with diverse sizes, shapes, and handling requirements, prompting more flexible and adaptive configurations. Simultaneously, scholars are emphasizing the integration of geometric innovation with operational strategies. While non-traditional layouts such as the Fishbone and Flying-V improve picker routing, they often remain decoupled from storage logic. The proposed CPU-based layout bridges this gap by combining spatial zoning with SKU turnover classification in a modular structure—introducing a hybrid logic akin to CPU cores, where distinct zones are optimized for task frequency, product variability, and overall system balance. In doing so, this work contributes a meso-level geometric design that embeds class-based slotting logic directly into the physical structure of the warehouse.

\section{Design Concept}
\label{sec:design_concept}
Modern warehouses face conflicting demands of throughput, efficiency, and space utilization. Inspired by heterogeneous computing architectures, the CPU layout introduces a spatially structured zoning model tailored to SKU turnover rates. Unlike traditional designs where layout and slotting policies are decoupled, the CPU layout integrates operational logic directly into its physical configuration. This approach enhances not only picker efficiency but also layout scalability and resilience to demand shifts.

\begin{figure}[h!]
    \centering
    \includegraphics[width=0.75\linewidth]{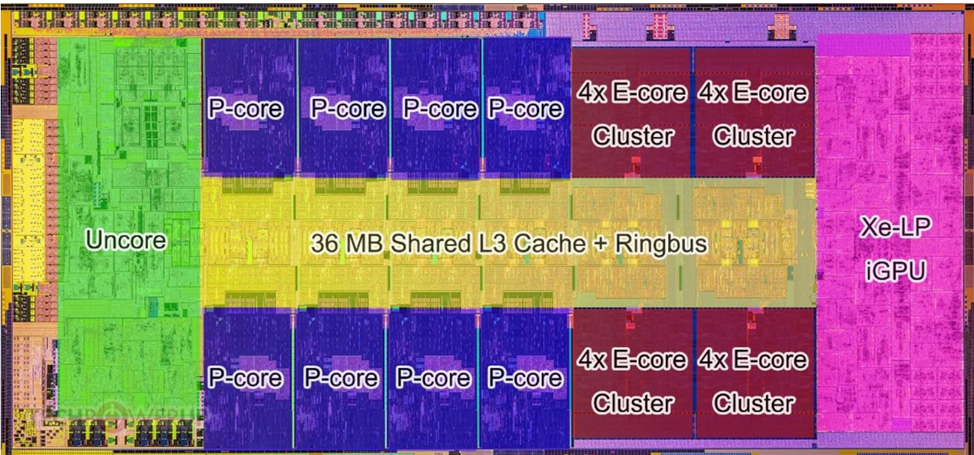}
    \includegraphics[width=0.75\linewidth]{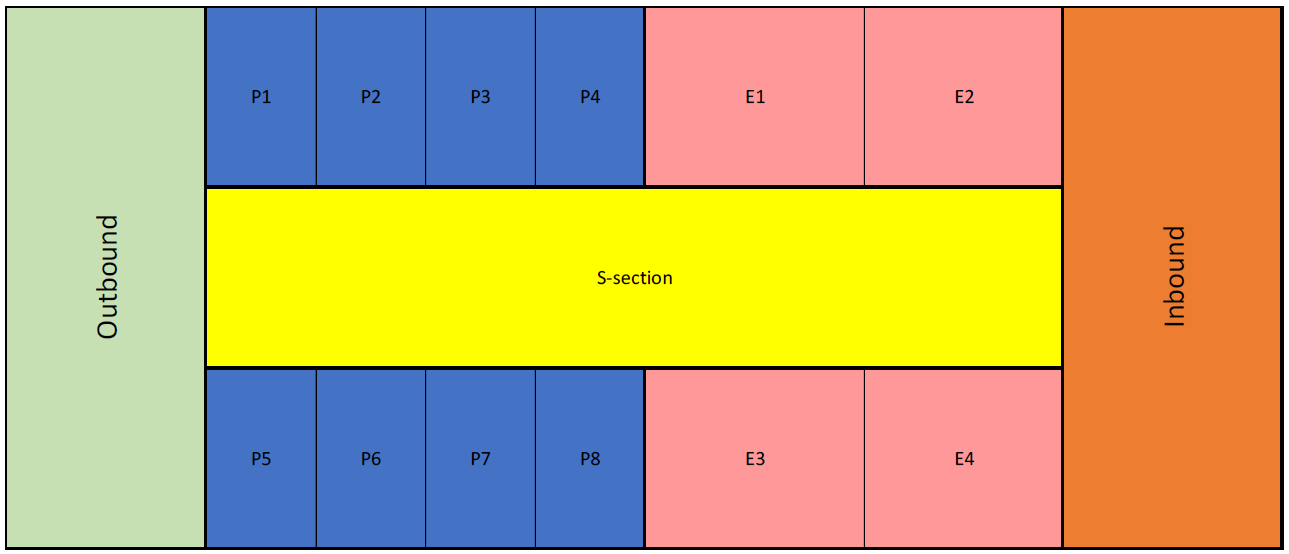}
    \caption{CPU microarchitecture-inspired warehouse layout (top: computing analogy; bottom: layout structure).}
    \label{fig:cpu-archi}
\end{figure}

Inspired by hybrid CPU architectures—specifically Intel’s 12th Gen Raptor Lake—the proposed warehouse layout applies a performance-efficiency-sharing (P/E/S) logic to physical space allocation. As shown in Figure~\ref{fig:cpu-archi}, modern processors integrate multiple specialized cores: 
\textit{Performance} (P) cores, optimized for high-intensity tasks; \textit{Efficiency} (E) cores, dedicated to routine or background processes; and \textit{Shared} (S) zones, representing cache regions that enable communication, coordination, and dynamic resource exchange among cores.

Analogously, our CPU-based warehouse layout segments space into:

\begin{itemize}
    \item \textbf{P-Zones}: Open-floor, high-accessibility areas designated for fast-moving SKUs. These prioritize pick speed and minimal congestion.
    \item \textbf{E-Zones}: High-density rack areas for medium-turnover SKUs, optimizing space usage while maintaining acceptable travel times.
    \item \textbf{S-Zones}: Flexible buffer areas that accommodate overflow, slow-movers, or dynamically reassigned SKUs to smooth workload imbalances.
\end{itemize}

This tri-zonal logic extends beyond conventional \textit{zone storage} and classical \textit{class-based storage}, both of which are often used in combination. In typical applications, zone storage segments the warehouse by handling characteristics (e.g., temperature zones or order types), while class-based storage ranks SKUs by demand frequency to improve picking efficiency. However, these approaches are usually layered without tightly integrating physical layout with access priority. The CPU layout, by contrast, structurally embeds class differentiation into the geometric design—assigning high-frequency SKUs to ground-level P-zones, low-frequency items to dense E-zones, and overflow to flexible S-zones. This results in a layout where storage logic and physical accessibility are co-designed, enhancing both speed and scalability.


\section{Simulation Methodology}
\label{sec:simulation_method}

The objective of the simulation is to evaluate the operational performance of the proposed CPU-based warehouse layout compared to conventional alternatives, under realistic order picking dynamics.

Discrete-event simulation (DES) is selected as the primary modeling approach due to its ability to capture complex, stochastic system behaviors over time. Unlike analytical or Monte Carlo methods, DES accommodates dynamic order flow, congestion, vehicle routing, and layout-specific constraints in a realistic manner.

The simulation model is parameterized using real-world operational data from a Dutch industrial warehouse, including SKU class frequencies (80/15/5 ABC split), pallet height distributions, and vehicle specifications. This grounding ensures the simulated outcomes are representative of actual warehouse performance.

A DMS model was developed in Siemens Tecnomatix Plant Simulation, a widely used industry platform for modeling logistics systems. The simulation focuses on the internal operations of a pallet warehouse: inbound unloading and storage, and outbound picking and dispatch. External production processes are excluded, but the outbound flow is synchronized to takt-based order waves. Layout-specific parameters such as aisle configuration, zone access, and SKU assignment policies are incorporated. The model simulates the layout alternatives under equivalent demand and labor inputs. The model simulates 80 operational days (excluding warm-up periods) using common random numbers for statistical control. All performance measures are aggregated across replications for robustness.

Two pallet formats are used—Euro (1200×800 mm) and Half Euro (800×600 mm)—with collar-based height distinctions affecting rack allocation and vehicle throughput. Pallets are categorized by collar height—ranging from 1 to 6 collars (200 mm to 1200 mm)—influencing both slotting and handling strategies.

The simulation reflects a high-volume manual pallet warehouse processing 2,000–2,100 pallets daily. Operations rely on reach trucks in narrow aisles, using double-deep racking over a footprint of 4,000–4,500 m\textsuperscript{2}. Inbound pallets arrive continuously; outbound orders are processed in timed waves. While the assumption of continuous inbound flow and batched outbound order release simplifies operational variability, it closely aligns with typical practices in warehouses serving manufacturing lines or fixed-time loading schedules. Empirical observations in partner facilities confirmed this pattern and informed the simulation input structure.
Operational constraints such as aisle width, turning radius, and stacking limits are modeled explicitly to maintain realism.
\subsection{Process Flow Logic}
\begin{figure}[h!]
    \centering
    \includegraphics[width=0.8\linewidth]{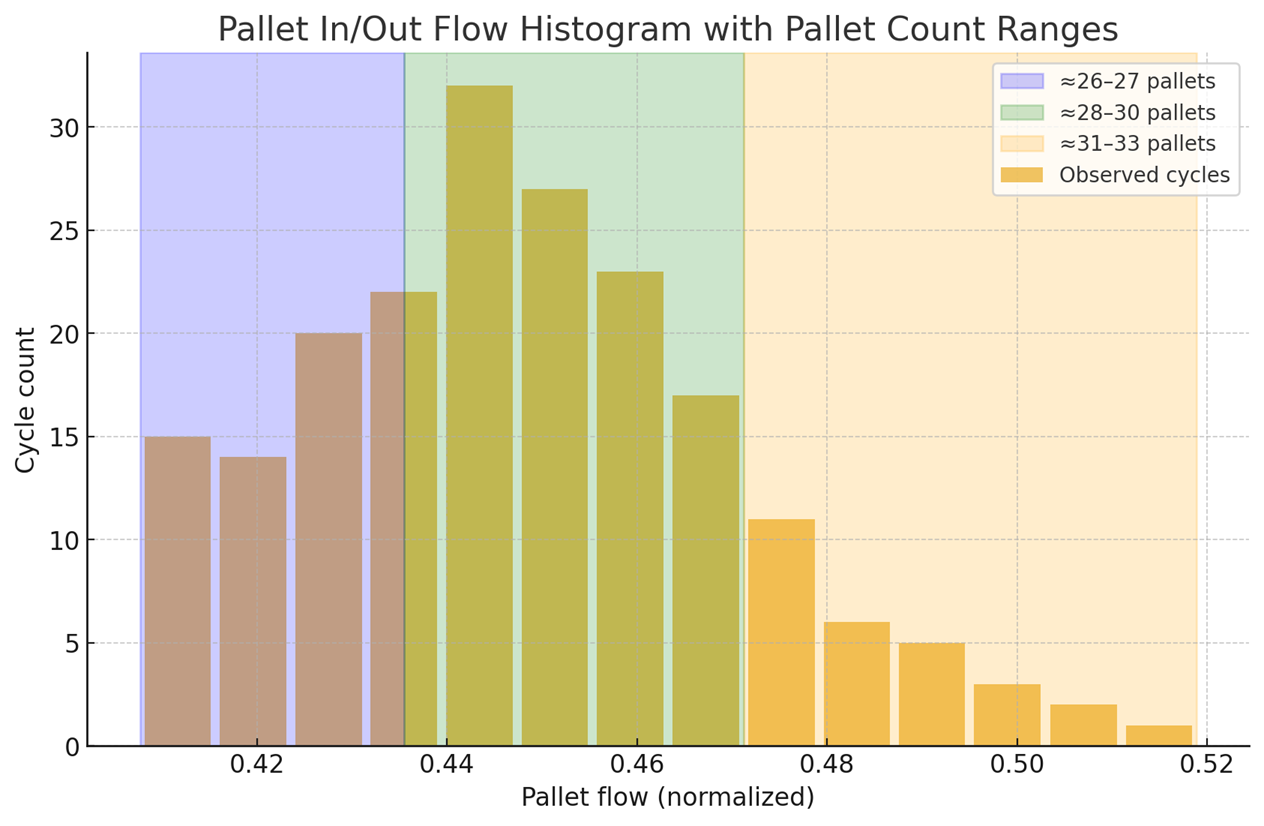}
    \caption{Pallet in/out flow histogram with cycle volume overlays. The shaded areas represent observed pallet count ranges per 25-minute cycle.}
    \label{fig:inout-hist}
\end{figure}
\begin{figure}[h!]
    \centering
    \includegraphics[width=0.8\linewidth]{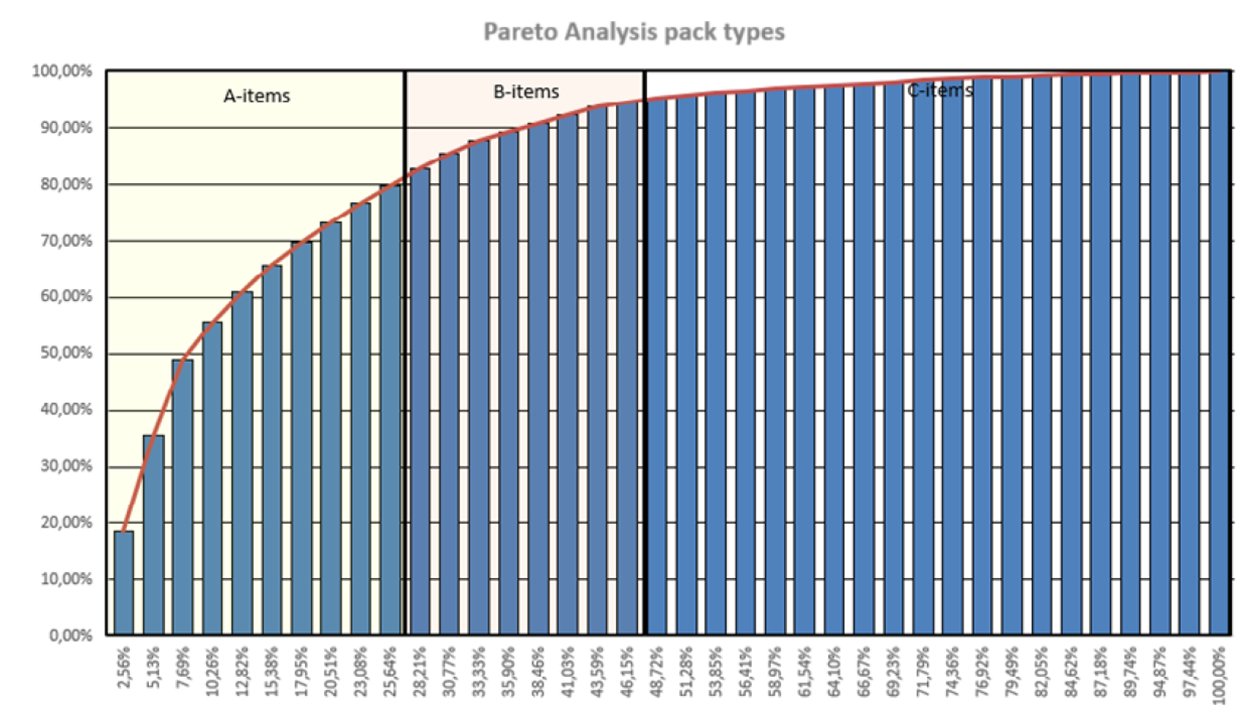}
    \caption{Pareto analysis of pack types based on demand share. A-items constitute 20\% of SKUs but 80\% of volume.}
    \label{fig:pareto-pack}
\end{figure}
\subsubsection{Inbound Process} Pallets arrive at the warehouse continuously and are modeled using a stochastic input generator with interarrival times drawn from a real-world log-normal distribution (see Figure~\ref{fig:inout-hist}). Upon generation, each pallet is assigned a demand classification—A, B, or C—based on an 80/15/5 distribution that mirrors typical production-driven fulfillment profiles (see Figure~\ref{fig:pareto-pack}). Additionally, a collar height (ranging from 1 to 6 collars) is assigned, which influences storage placement depending on the applicable layout policy.

Inbound pallets are unloaded by reach trucks and routed to their assigned storage slots. Each vehicle can handle up to 9 collar units per trip. Pallet handling involves vertical fork adjustments and horizontal transport time, both of which are explicitly modeled using 3D kinematic constraints and layout-specific pathfinding logic.

Routing decisions are governed by dynamic shortest-path algorithms that account for aisle width constraints and vehicle maneuverability, including minimum turning circles. Once a pallet is delivered, the reach truck either proceeds to the next assigned destination or returns to the staging area based on the current task queue.

\begin{figure}
    \centering   \includegraphics[width=0.8\linewidth]{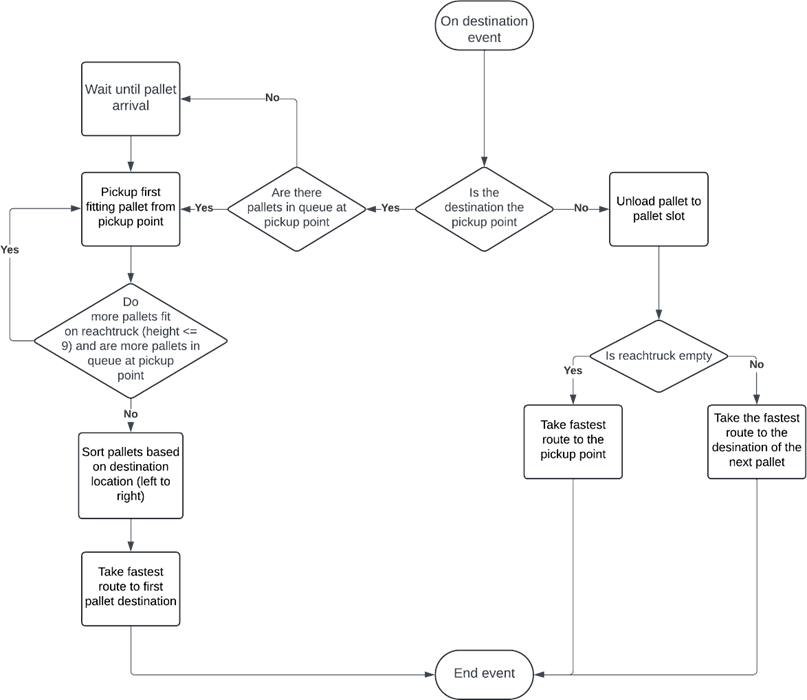}
\caption{Inbound Reach Truck Handling Logic and Routing Flowchart}
    \label{fig:inbound-flowchart}
\end{figure}

The complete behavioral logic of the inbound process is visualized in Figure~\ref{fig:inbound-flowchart}, which maps key decision points such as queue checks, load consolidation, destination sorting, and optimal route calculation. This diagram serves as a high-level abstraction of the vehicle logic implemented in the simulation.

\subsubsection{Outbound Process}

Outbound operations are initiated every 25 minutes, synchronized with the takt time of the production system. Each outbound order comprises multiple SKUs, with the number of required pallets per SKU following a normal distribution to reflect typical production-driven variability. Pallet selection is executed dynamically, based on the current warehouse inventory and weighted by SKU picking frequency to simulate realistic ABC demand patterns.

Dedicated outbound reach trucks—functionally distinct from inbound vehicles—are responsible for retrieving pallets and transporting them to outbound staging zones. Upon assignment, each truck calculates the shortest admissible route to the next target pallet, incorporating both horizontal and vertical movements to access multi-level racking. 

\begin{figure}
    \centering
    \includegraphics[width=\linewidth]{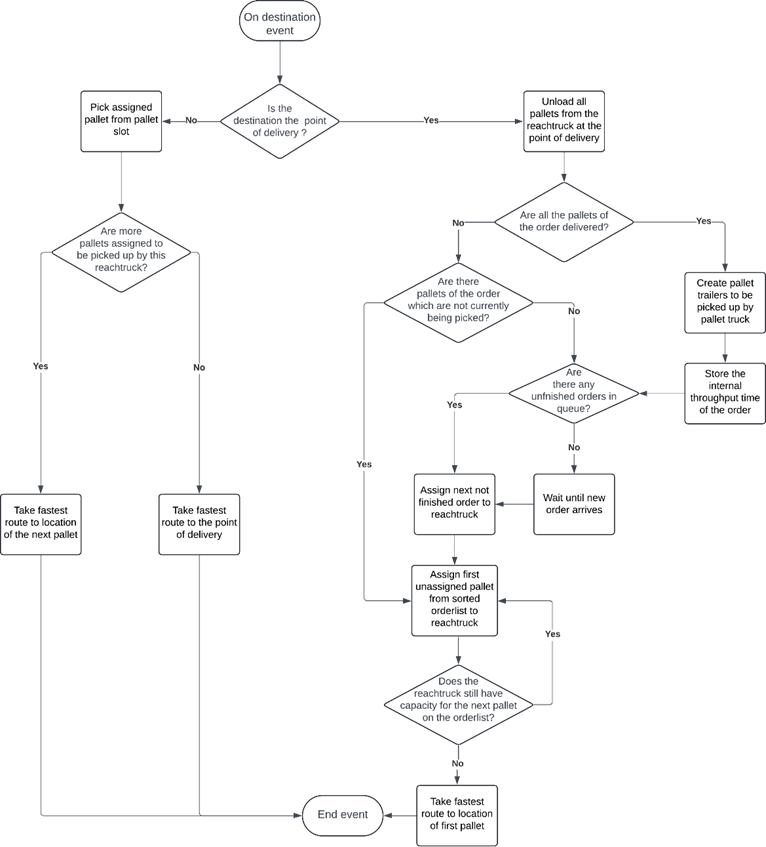}
    \caption{Outbound pallet retrieval logic — flow diagram of dispatch handling, order assignment, and delivery sequencing.}
    \label{fig:outbound-flowchart}
\end{figure}

Once a pallet is delivered, the truck evaluates whether additional pallets from the same order remain within its capacity and proximity. If so, it proceeds to retrieve the next item; otherwise, it returns to the staging area or awaits the next order release. This logic ensures efficient sequencing and reduces travel redundancy. The entire decision flow of outbound pallet handling is visualized in Figure~\ref{fig:outbound-flowchart}, illustrating assignment, order progression, and delivery confirmation.

\subsubsection{Routing and Task Sequencing} All pallet movements are governed by a shortest-path routing algorithm. For inbound operations, pallets are sequenced based on left-to-right rack positioning to minimize search and travel delays. For outbound retrievals, the route is dynamically recalculated after each delivery to maintain efficiency.

\subsection{Key Performance Indicators (KPIs)}

The following four KPIs are used to assess layout efficiency:

\begin{itemize}
    \item \textbf{Throughput Time (minutes):} Average time from order release to pallet readiness for outbound dispatch. This reflects internal process velocity.

    \item \textbf{Required Full-Time Equivalents (FTE):} Estimated number of full-time operators (reach truck drivers) needed per shift. Calculated as average simultaneous usage multiplied by two.

    \item \textbf{On-Time Order Completion (\%):} Percentage of outbound orders fulfilled within the target takt time of 25 minutes, reflecting delivery punctuality.

    \item \textbf{Required Warehouse Area (m\textsuperscript{2}):} Total floor space dedicated to the storage layout, based on actual aisle/rack dimensions. Staging and buffer zones are excluded.
\end{itemize}

\section{Results}
\subsection{Visualized Warehouse Layouts for Comparison}
\begin{figure}[h!]
    \centering
    \includegraphics[width=0.7\linewidth]{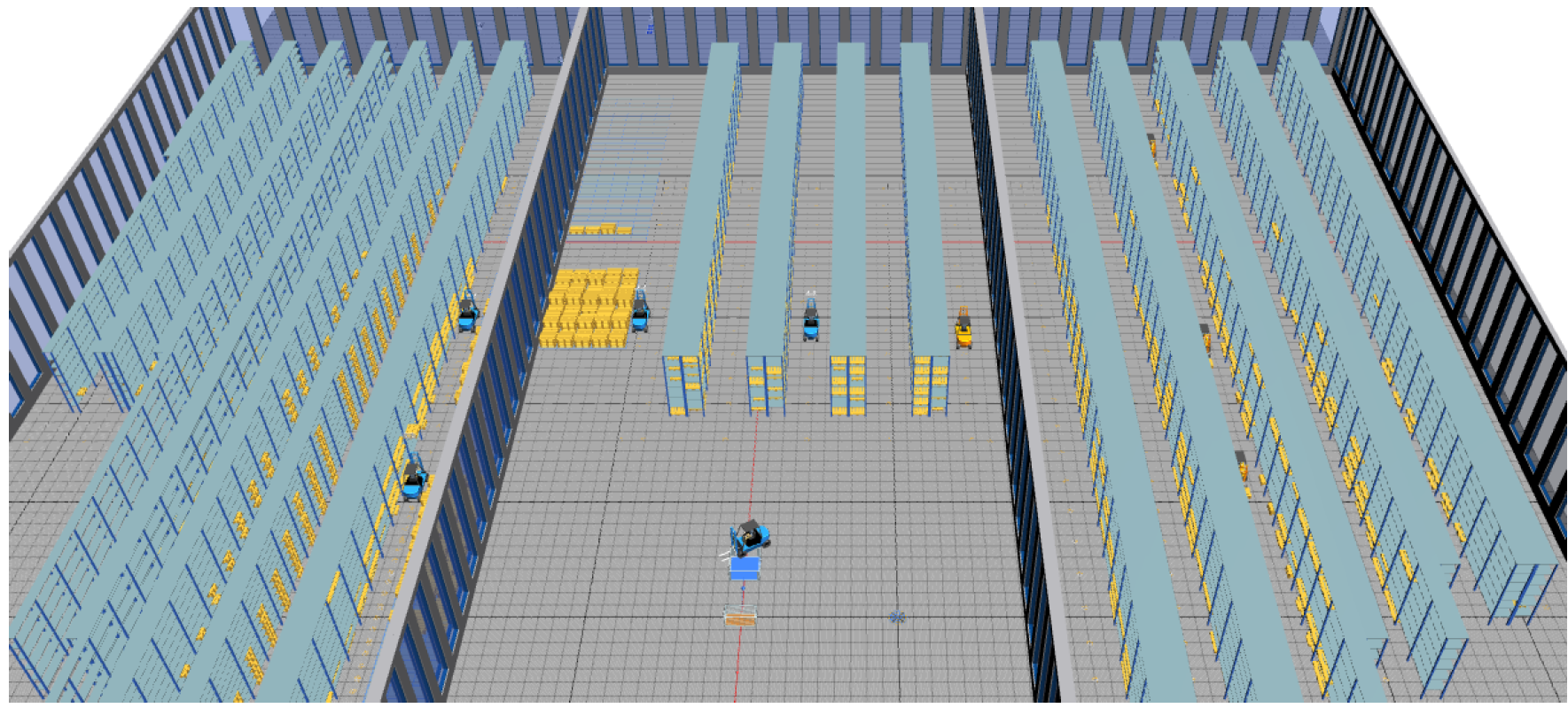}
    \caption{Current warehouse layout in simulation model. Blue vehicles represent outbound reach trucks; orange represent inbound.}
    \label{fig:current-layout}
\end{figure}
We show in this subsection the visual representation of different layouts generated in the simulation. Figure~\ref{fig:current-layout} illustrates the current warehouse configuration, which serves as the baseline for comparative analysis. The current warehouse layout, shown in Figure~\ref{fig:current-layout}, is a conventional double-deep racking system organized in a rectangular grid with straight, parallel aisles. Both inbound and outbound staging zones are positioned along the bottom edge of the layout, enabling consolidated docking operations. While the layout does not employ formal zoning, it incorporates a basic class-based logic by storing fast-moving SKUs (A-items) on the ground floor to reduce retrieval time.

\begin{figure}[h!]
    \centering  \includegraphics[width=0.8\linewidth]{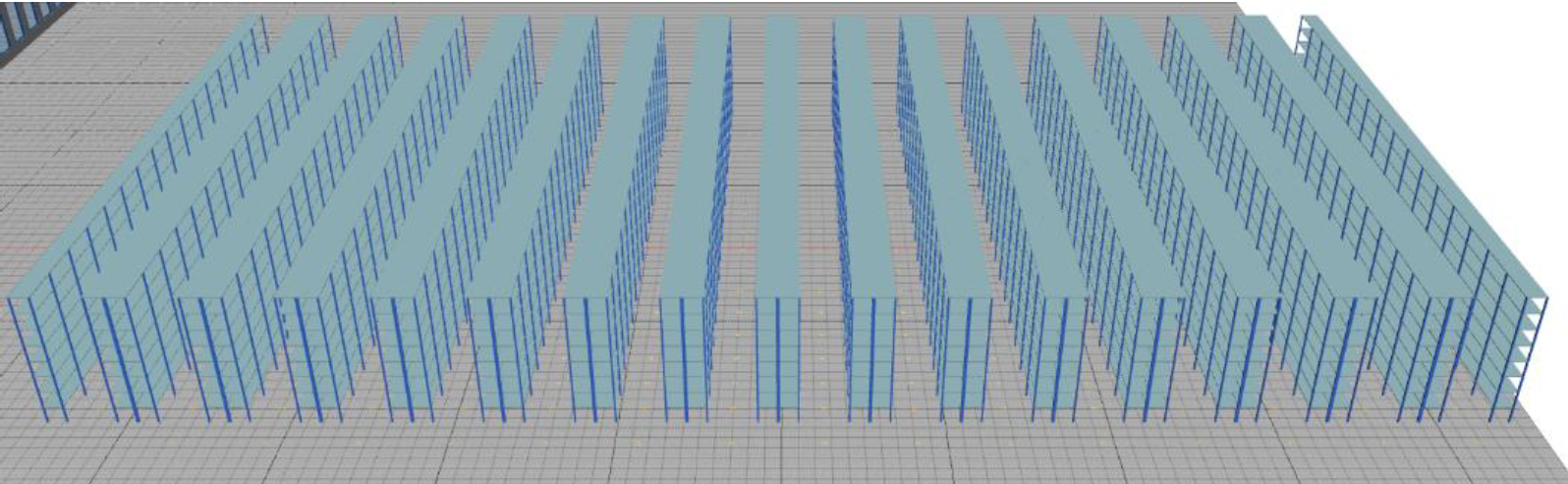}
    \caption{Visual representation of conventional layout.}
    \label{fig:c-layout}
\end{figure}

\begin{figure}[h!]
    \centering
    \includegraphics[width=0.8\linewidth]{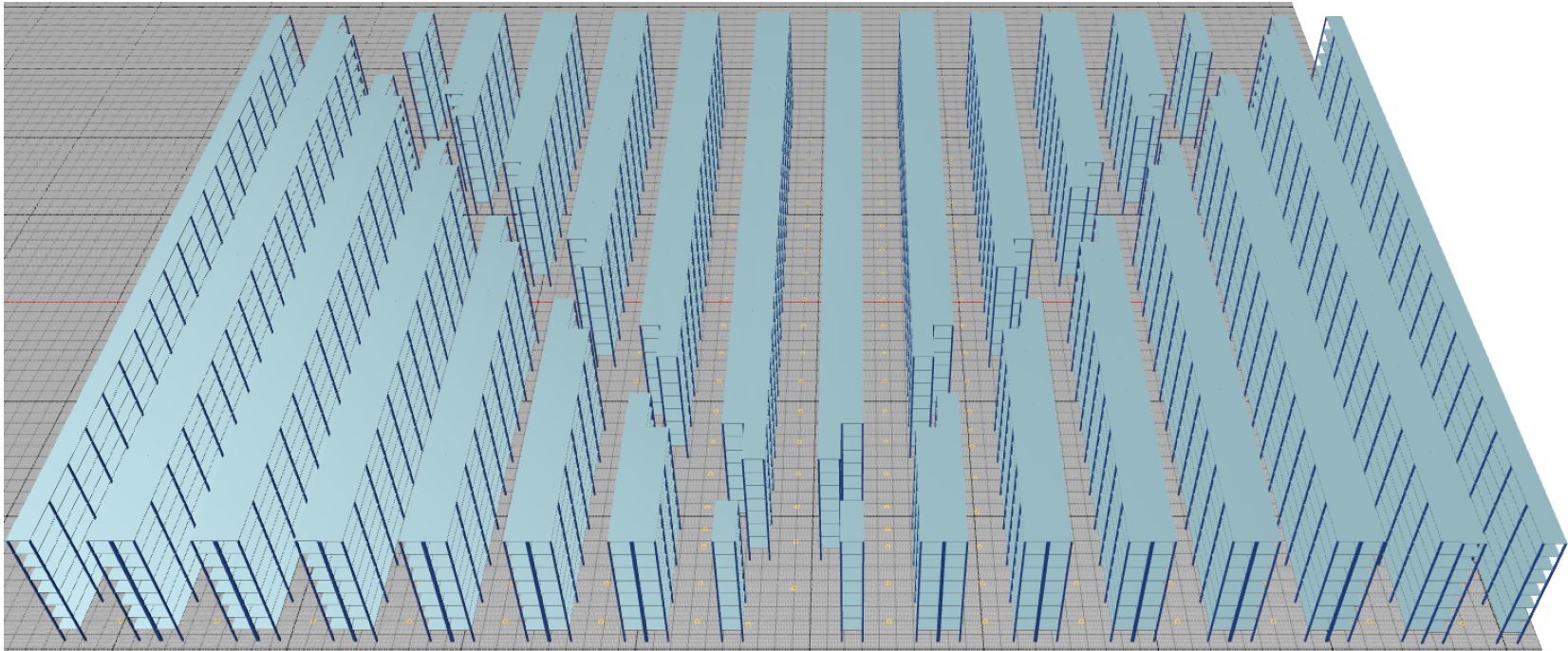}
    \caption{Visual representation of Flying-V layout.}
    \label{fig:fish_v}
\end{figure}

\begin{figure}[h!]
    \centering
    \includegraphics[width=0.8\linewidth]{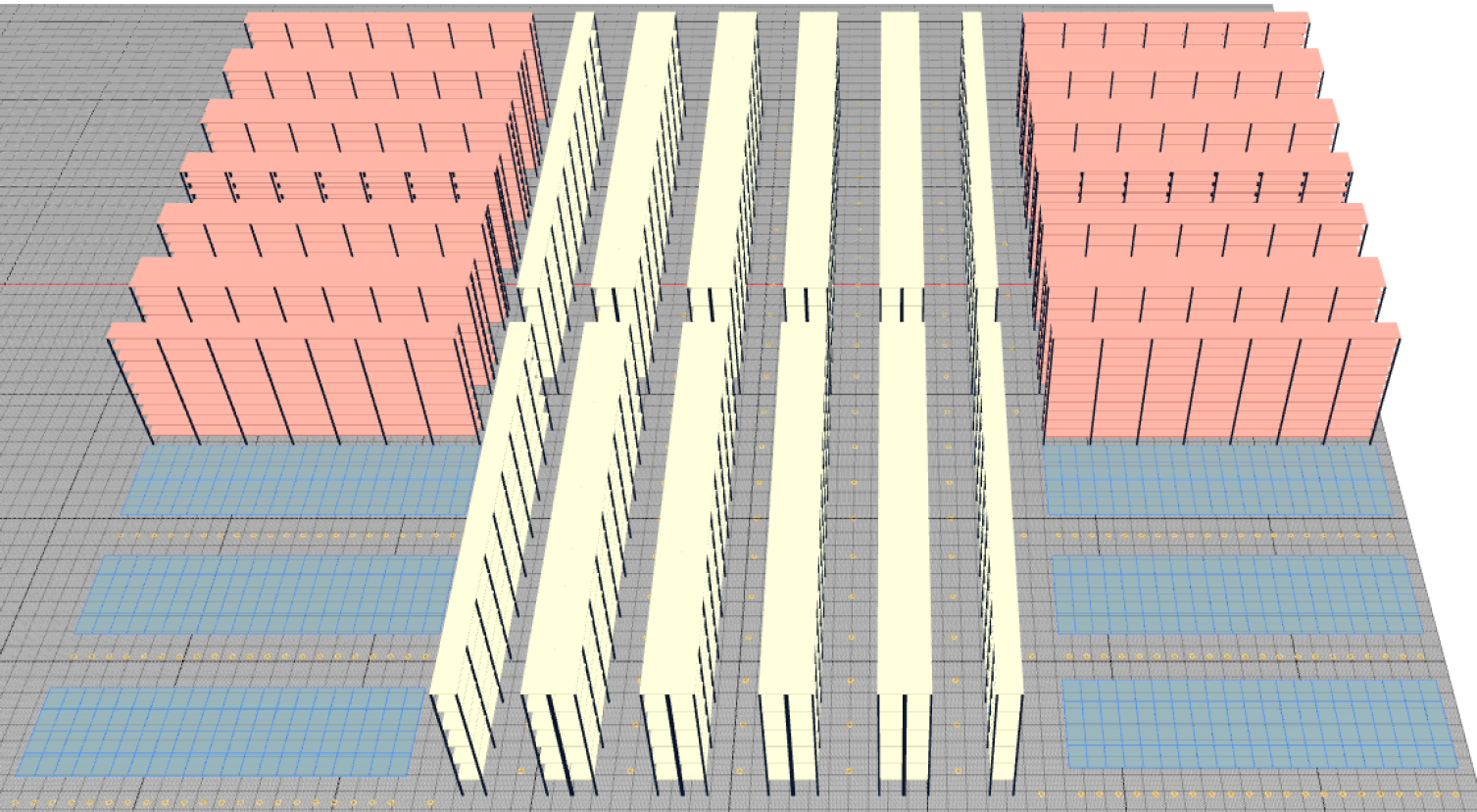}
        \caption{Simulated CPU-based warehouse layout showing zone assignment.}

    \label{fig:cpu-layout}
\end{figure}

Figures~\ref{fig:c-layout} through \ref{fig:cpu-layout} illustrate the geometric progression from conventional orthogonal layouts to Flying-V and CPU-based configurations. While the figures emphasize internal rack structures and aisle orientation, the inbound and outbound staging zones are positioned consistent with the real warehouse layout—located along the bottom edge—but are not explicitly visualized in the diagrams.

The Flying-V layout aligns aisles diagonally towards outbound areas, optimizing picker travel paths (see Figure~\ref{fig:fish_v}). Although this layout achieves reduced travel distances, it typically requires more area due to its non-linear aisle configuration and can introduce complexity in warehouse operations.

\subsection{Comparison of Storage Policies and Layouts}

The layouts evaluated in this study differ not only in physical structure but also in storage policies. The CPU-based layout employs a structured, zone-specific class-based storage policy (see Figure~\ref{fig:cpu-layout}). A-class SKUs (high-frequency) are assigned to P-Zones (in color blue) for fast access and minimal picker travel. C-class SKUs (low-frequency) are stored in high-density E-Zones (in color orange) to maximize space utilization. B-class and overflow SKUs are allocated to flexible S-Zones (in color yellow) to balance load and adapt to demand fluctuations. Pallet height is also considered—taller pallets are excluded from ground-level P-Zones due to clearance constraints and picking ergonomics, and are instead routed to E-Zones or upper levels of S-Zones where vertical space is more effectively utilized.

In contrast, the \textbf{Conv.–Random} layout allocates SKUs without consideration of demand frequency, resulting in inefficient retrieval times—especially for fast-moving items. The \textbf{Conv.–Class} layout applies a class-based policy but lacks the dedicated zoning of the CPU design, instead concentrating A-class items near outbound areas within a flat structure. \textbf{Flying-V (ABC)} maintains the same ABC storage logic but improves picker routing through aisle reorientation towards the outbound zone.

\subsubsection{Storage and Retrieval Logic by Layout Type}

The CPU layout integrates class-based slotting directly into its spatial structure. High-frequency SKUs are assigned to low-lift, ground-level P-zones to minimize travel time. Medium-frequency items are routed to flexible S-zones, while low-frequency SKUs are stored in dense, high-rack E-zones. This zoning mirrors CPU core logic and embeds prioritization into the layout itself.

In contrast, conventional class-based layouts apply ABC slotting heuristically—placing A-items closer to outbound zones—without modifying geometric design. Random storage layouts ignore demand frequency entirely, focusing only on space availability. Flying-V layouts improve routing geometrically, but only when combined with class-based logic do they approach CPU-like efficiency. Table~\ref{tab:storage-policies} summarizes the storage logic across layouts.

\begin{table*}[th]
\centering
\caption{Storage and Retrieval Logic Across Layout Types}
\label{tab:storage-policies}
\begin{tabularx}{\textwidth}{|l|X|}
\hline
\textbf{Layout Type} & \textbf{Storage and Retrieval Logic} \\
\hline
CPU-Based & Zone-based class storage: A-items in ground-level P-zones for fast access; C-items in dense E-zones; B-items in flexible S-zones. Aligns spatial zones with turnover rates. \\
\hline
Conv.–Random & No class consideration. Pallets stored in first-available slots. High utilization but poor picker efficiency due to randomized retrieval paths. \\
\hline
Conv.–Class-Based & Applies ABC classification by distance only. A-items near outbound zone, others further away. No zoning structure; moderately improved efficiency. \\
\hline
Flying-V (Random) & Diagonal aisle geometry with randomized SKU placement. Routing is geometrically efficient, but retrieval remains stochastic. \\
\hline
Flying-V (ABC) & Aisles diagonally directed to outbound. A-items located centrally for quick access; B- and C-items placed outward. Combines routing and slotting optimization. \\
\hline
\end{tabularx}
\end{table*}

\subsection{Simulation results}
The simulation results are summarized in Table~\ref{tab:results}. The comparison focuses on four representative configurations: \textbf{Conv.–Random}, \textbf{Conv.–Class}, \textbf{Flying-V (Random)}, \textbf{Flying-V (ABC)}, and the proposed \textbf{CPU} layout. These were selected to represent a variety of storage policies and spatial designs.

\begin{table}[h!]
\centering
\caption{Performance Comparison of Layouts — throughput time in minutes, labor in FTEs, and area in square meters}

\label{tab:results}
\begin{tabular}{|l|c|c|c|c|}
\hline
Layout & Throughput Time (min:sec) & FTE & On-Time (\%) & Area (m$^2$) \\
\hline
Current & 16.25 & 26 & 98.82 & 7249\\
Conv.–Random & 16:08 & 20 & 100 & 4023 \\
Conv.–Class & 15:53 & 20 & 100 & 4023 \\
Flying-V (Random) & 16:03 & 20 & 100 & 4477 \\
Flying-V (ABC) & 15:49 & 20 & 100 & 4477 \\
CPU & 14:56 & 20 & 100 & 4112\\
\hline
\end{tabular}
\end{table}

All tested layouts deliver performance improvements across key KPIs compared to the existing warehouse configuration. The use of ABC classification in the Conv.–Class layout significantly reduces picker travel time relative to Conv.–Random. The Flying-V layout offers further gains by orienting aisles toward the outbound zone, improving routing efficiency. Among the alternatives, the CPU layout achieves the shortest throughput time and maintains high on-time performance while using slightly more space.

Although the CPU layout requires a modest increase in floor area (+2.2\% compared to the conventional layout), this trade-off is justified by its significant improvement in throughput efficiency. Throughput time—defined as the interval from order release to pallet consolidation—is a critical performance driver in high-volume warehouses. Even small reductions at the order level can translate into substantial time and labor savings. The CPU layout shortens average throughput time by approximately 1–2 minutes compared to other configurations, resulting in several hours of operational gain per day. It also lowers buffer requirements, reduces cycle time variability, and minimizes late-order risk. Thus, despite its slightly larger footprint, the CPU layout offers strong advantages in temporal efficiency and process stability.

\section{Discussion}
\begin{figure}
    \centering
    \includegraphics[width=0.75\linewidth]{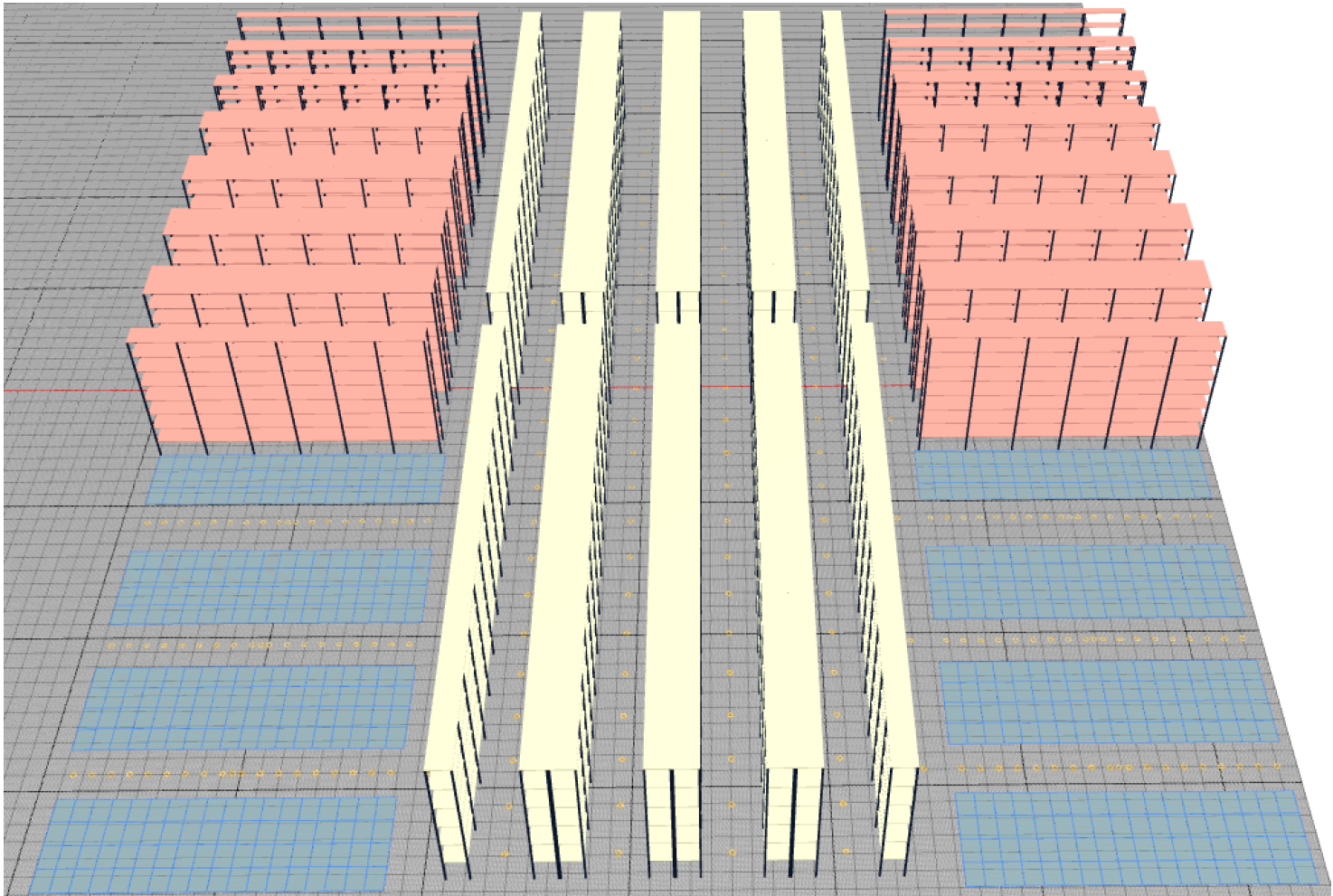}
    \caption{Visualization of the changed areas of CPU-based layout.}
    \label{fig:cpu-layout-2}
\end{figure}

The results illustrate clear advantages of CPU-based layouts. The CPU layout achieves superior picker efficiency through optimal P-Zone placement, despite a slightly larger area requirement. It provides a balanced approach, improving space efficiency and throughput, suitable for environments with spatial constraints. This research is the first attempt to see the performance of this layout. How to find out the best size of different zones is not discussed here. But we make the a slit different layout illustrated in Figure~\ref{fig:cpu-layout-2} to store the same number of pallets but with larger P-zones. We can get already even better throughput time, namely 14:40 minutes. 


Limitations of our simulation include simplifications in picker behavior and congestion. The model assumes ideal travel speeds and no delays due to real-time obstruction or equipment failure. Pallet load diversity and replenishment strategies are also abstracted. Despite this, the model captures relative layout performance reliably under stable order conditions.

\section{Conclusion}
This study presents the CPU-based layout as a novel integration of zoning and class-based storage in manual pallet warehouses. Compared to conventional and Flying-V layouts, the CPU layout achieves notable improvements in throughput time and on-time order fulfillment, with only marginal increases in required area.

From a practical standpoint, the CPU concept supports modular planning, scalable zone assignment, and streamlined picker routing—making it well-suited for high-throughput, takt-driven environments. Practitioners may particularly benefit from the ability to tailor storage zones to SKU turnover dynamics and traffic behavior.

Future research should examine adaptive zone sizing mechanisms that respond to shifting SKU profiles or seasonal demand patterns. Additionally, assessing CPU-based logic under alternative slotting policies—including randomized or dynamically assigned SKU storage—could reveal its robustness in volatile environments. The modular nature of CPU zones also invites exploration into multi-story or automated robotic applications, extending its utility to high-density or vertical fulfillment centers.

\bibliographystyle{splncs04}
\bibliography{references}

\end{document}